\documentstyle[12pt,aaspp4,psfig]{article}    
\def\ergs   {~erg~s$^{-1}$}

\def\gax    {${_>\atop^{\sim}}$}
\def\etal	{et~al.}

\def\cm2   {~cm$^{-2}$}

\begin{document}

\title{	Two Neutron Star SXT in Quiescence: \\
		4U~2129+47 and EXO~0748-676} 
\author{Michael R. Garcia\altaffilmark{1} and
Paul J. Callanan\altaffilmark{2}}

\altaffiltext{1}{Harvard-Smithsonian Center for Astrophysics, MS-4, 60
Garden St., Cambridge, MA 02138; email: mgarcia@cfa.harvard.edu}
\altaffiltext{2}{Physics Department, University College, Cork,
Ireland; email: paulc@ucc.ie}

\begin{abstract}

We report on x-ray observations of two soft x-ray transients
containing neutron stars, 4U~2129+47 and EXO~0748-676.  Our emphasis is
on the quiescent properties of these sources.

The x-ray spectrum and lightcurve of the eclipsing Low-Mass X-ray
Binary (LMXB) 4U~2129+47 was measured with the ROSAT PSPC during its
current quiescent state.  The quiescent x-ray luminosity of $\sim
10^{32.7}$\ergs\/ and blackbody temperature of ${\rm kT} \sim 0.21$~keV are
similar to other quiescent LMXB.  The quiescent x-ray light curve
appears to show orbital modulation, but the statistics are
insufficient to distinguish between a v-shaped partial eclipse (as
seen in the high state) or a total, square wave eclipse.  The
similarity in the luminosity and temperature to other (non-eclipsing)
quiescent LMXB implies that the vertical structure in the disk which
blocked our direct view of the neutron star in the high state has
collapsed, and the neutron star is seen directly.

EXO~0748-676 was serendipitously observed with the Einstein IPC in
quiescence before it was discovered as a bright transient with EXOSAT.
Our re-analysis of this quiescent observation finds ${\rm L_x} \sim
10^{34.0}$\ergs, and blackbody temperature of ${\rm kT} \sim 0.22$~keV, again
similar to other LMXBs in quiescence.

\end{abstract}


\section{Introduction:}

4U~2129+47 and EXO~0748-676 are two low-mass x-ray binaries which
undergo high/low transitions in their X-ray flux.  Both are viewed
nearly edge on, and their eclipse lightcurves show evidence for an
extended x-ray emission region often called an accretion disk corona
(ADC).  4U~2129+47 is currently the only ADC source in a low state,
and was very well studied in the 1980s when it was in the high state.
EXO~0748-676 is currently in the high state, but was serendipitously
observed in a low state by the Einstein Observatory before it was
discovered as a bright transient source (Parmar \etal\/ 1986).

The optical and x-ray lightcurves of 4U~2129+47 in the 1970s and 1980s
showed a v-shaped partial eclipse, which lead to the model of a
partial eclipse of an extended x-ray emission region (McClintock
\etal\/ 1982).  The inclination is believed to be high enough so that
the accreting neutron star is not directly visible, but is shielded
from our view by vertical structure at the outer edge of the accretion
disk.  The x-rays we do observe are only the few percent of those
emitted from the vicinity of the neutron star which are scattered into
our line of sight by the ADC.  EXOSAT observations in 1983 failed to
detect the source, and the optical modulation seen previously was also
found to be missing (Peitch \etal\/ 1986).  We previously reported on
observations of this source in quiescence with the ROSAT HRI, which,
although sufficient to detect the source at $L_x \sim 10^{33.5}$erg/s,
could not measure the spectrum or a detailed lightcurve (Garcia 1994).

EXO~0748-676 was discovered in outburst with EXOSAT (Parmar \etal\/ 1986).
While it shows sharp (square wave) eclipses, there is a residual flux
of a few percent at the bottom of the eclipse.  This residual flux is
interpreted as due to an accretion disk corona, which covers a 
large geometric area and is therefore not fully eclipsed. 
It was serendipitously observed with the Einstein IPC 
before it entered the high, discovery state.  

Because of their transient nature, 4U~2129+47 and EXO~0748-676 are
often referred to as soft x-ray transients (SXT).  Several SXT have
been shown to have mass functions $>3M_\odot$, and on this basis are
considered black holes, or BH~SXT (van Paradijs and McClintock 1995,
Tanaka and Shibazaki 1996).  Because they show Type~I x-ray bursts,
4U~2129+47 and EXO~0748-676 clearly contain neutron stars (NS), and
are therefore NS~SXT.  Observations of BH~SXT provide good evidence
that they accrete via advection dominated accretion flows (ADAF,
Narayan, McClintock \& Yi 1996, Narayan, Barret \& McClintock 1997),
rather than by thin disks, when they are in quiescence.  The
theoretical basis for ADAF applies equally well to NS~SXT in
quiescence (Yi \etal\/ 1996).  Detailed study of the accretion
including the effects of a rotating magnetic field certainly allow
that ADAFs exist in NS SXT (Menou \etal\/ 1999).  
Comparison of outburst and quiescent x-ray luminosities in SXT may
have provided evidence for the existence of event horizons in BH~SXT. 
(Narayan, Garcia, \& McClintock 1997a, Garcia \etal\/ 1998, Asia
\etal\/ 1998). Extending the small sample of objects used in these
studies may help to prove (or disprove) this fundamental result; hence
these observations of two NS~SXT in quiescence are of interest. 

In addition, these source are of interest because they offer
opportunity to study the way the structure of the accretion disk and
ADC are effected by wide variations in the x-ray flux of the central
source.  Accretion disk models predict that the both the vertical
structure and the run of temperature with radius changes dramatically
under the influence of strong x-ray irradiation (Shakura \& Sunyaev
1973, Vrtilek \etal\/ 1990), but there are few chances to confirm
these predictions over wide variations in luminosity.  The ADC is formed
as a result of the strong x-ray irradiation (Begelman, McKee \& 
Shields 1983), and once again there are few opportunities to study its
structure (in a single source) over very wide ranges in x-ray
luminosity.

\section{Observations:}

EXO~0748-676 was serendipitously observed in its low state during a
10~ks observation with the {\sl Einstein}\/ Observatory in May~1980.
The ROSAT PSPC observed 4U~2129+47 in the low state for a total of
$\sim$30~ks on 3~June~1994, and these data were processed with the
standard SASS pipeline.  Both datasets were analyzed using IRAF/PROS
V2.3.1 and XSPEC V10.0.  Poisson weighting of the errors (Gehrels 1986)
was used.  The results from PROS and XSPEC were found to be
consistent.  When necessary, we grouped the data in larger energy
bins in order to maintain the number of counts per bin \gax~9.
Previous studies have shown that under these conditions, the results
from Poisson weighting of $\chi^2$ fitting are consistent with those
from maximum likelyhood methods (Narayan, Barret \& McClintock 1997b)

In order to guide our extraction of the source counts from the images,
we first generated the azimuthally averaged radial profile of counts
centered on the sources.  For 4U2129+47 we found that a $0.5'$ radius
extraction maximized the S/N.  For EXO~0748-676, we found that a $2'$
radius extraction maximized the S/N.  For both sources the background
was determined from a larger annulus around the source.

For both sources, we found that a variety of simple models
(Raymond-Smith thermal, bremsstrahlung, blackbody, power law) gave
equally acceptable fit results.  As our main interest in the spectral
fitting is to derive quiescent luminosities, in order to allow
comparisons between NS and BH luminosity swings (i.e., 
Narayan \etal\/ 1997a, Asai \etal\/ 1998) 
we limit ourselves below to
simple black-body fits and compute unabsorbed luminosities over the
0.5-10~keV range.  We note that the computed luminosities are
insensitive to the exact form of the spectrum, for example, using a
bremsstrahlung spectrum results in only 10\% changes in the
luminosities. 

\subsection{EXO~0748-676 Quiescent Spectrum}

The $\sim 70$~source counts extracted from the IPC image were fit to a
variety of simple spectral models.  The best fit blackbody parameters
are ${\rm kT} = 0.14$~keV, ${\rm log(N_H)} = 21.9~$cm$^{-2}$.  The very low
number of counts allow a wide range in acceptable parameters.  In
order to limit the parameter space, we restrict ourselves to log(${\rm
N_H}) = 21.35$~cm$^{-2}$ (which is within the allowed range).  This is
the value predicted by the relation of Predehl and Schmitt (1995),
given the optical reddening of E(B-V)$=0.42 \pm 0.03$ (Schoembs and
Soeschinger 1990).  We then computed the confidence regions for
temperature and emitted (unabsorbed) luminosity, as shown in Figure~1.
The best fit blackbody temperature and 0.5-10.0~keV luminosity,
assuming a distance of 10~kpc (Parmar \etal\/ 1986), are

$${\rm kT = 0.22^{+0.14}_{-0.10}~keV}$$

$${\rm L_x} = 1.0^{+0.5}_{-0.2} \times 10^{34} {\rm ergs~s^{-1}}$$

These results are fairly insensitive to the value of ${\rm N_H}$
assumed, in that a 50\% increase results in only a 15\% decrease in
best fit temperature and a 20\% increase in the emitted 0.5-10.0~keV
luminosity.  This energy band contains \gax~70\% of the bolometric
luminosity at the best fit temperature.  At the 
lowest temperatures allowed by the 90\% confidence interval
(Figure~1), this band contains only 30\% of the bolometric luminosity.

The effective radius of this possible blackbody emitter (${\rm R = (
L_{bol}/4\pi\sigma T^4)^{1/2}}$ ) is ${\rm R} = 8^{+12}_{-5}$~km, comparable
to the radius of a neutron star.

\subsubsection{4U~2129+47 Quiescent Spectrum}

Various simple models (as above) give acceptably good fits to the
$\sim 200$ source counts extracted from the image, and the fit
statistics show no preference for one model over any other.  The best
fitting black body model has ${\rm kT} = 0.18$~keV and ${\rm log(N_H)} =
21.5~$cm$^{-2}$.  As with EXO~0748-676, we fix the absorption at the
optically determined value (which is within the fit range) in order to
reduce the allowed parameter range.  The optical extinction has been
found to be $A_V = 0.9$ (Cowley \& Schmidtke 1990), which corresponds
to ${\rm log(N_H)} = 21.2$ (Predehl \& Schmitt 1995).  We then
computed the confidence regions for temperature and emitted
(unabsorbed) luminosity, as shown in Figure~2.  The best fit blackbody
temperature and 0.5-10.0~keV luminosity, assuming a distance of
6.3~kpc (Parmar \etal\/ 1986) are

$${\rm kT = 0.21_{-0.03}^{+0.04} keV}$$

$${\rm L_x = 5.3^{+0.8}_{-1.0} \times 10^{32} ergs~s^{-1} }$$

We note that this ${\rm A_V}$ is measured to the F7IV counterpart of
4U2129+47, which is clearly not the secondary transferring mass to the
neutron star.  In using this ${\rm A_V}$ and a spectroscopic distance of
6.3~kpc (Cowley \& Schmidtke 1990) we are implicitly assuming that
this star is in the physical proximity of the mass transferring binary
(Thorstensen \etal\/ 1988, Garcia \etal\/ 1989, Cowley \& Schmidtke
1990, van Paradijs \& McClintock 1995.)

Once again, these results are fairly insensitive to the value of ${\rm
N_H}$ assumed, in that a 50\% increase results in only a 10\% drop in
best fit temperature and a 40\% increase in the emitted bolometric
luminosity.  The 0.5-10.0~keV band contains $\sim 80$\% of the
bolometric luminosity at the best fit temperature, and $\sim 70$\% at
the lowest allowed temperature.

The effective radius of this possible blackbody emitter (${\rm R =
(L_{bol}/4\pi\sigma T^4)^{1/2}}$ ) is $R = 1.7^{+0.5}_{-0.6}$~km,
substantially smaller than the radius of a neutron star.

We previously reported the quiescent ROSAT HRI flux (Garcia 1994),
assuming the spectral parameters measured in the high state.  This
overestimates the flux by a factor of two.  When the softer quiescent
spectrum determined above is used we calculate an emitted luminosity
(0.5-10.0~keV) of $9.7 \times 10^{32}$ergs~s$^{-1}$.  The 68\%
confidence bounds on the PSPC spectrum correspond to a $\sim 20\%$
uncertainty in the calculation of the HRI flux.  Thus it appears that
the source has faded by a factor of $\sim 2$ during the 2.5~year
interval between the HRI and PSPC observations.

\subsubsection{4U~2129+47 Quiescent Lightcurve}

We generated the quiescent lightcurve of 4U~2129+47 (Figure~3) by
binning the background subtracted PSPC data into 7 bins based on the
McClintock \etal\ 1982 ephemeris.  Two other lightcurves are plotted
in Figure~3: The scaled on-state lightcurve re-extracted from the IPC
CD-ROM archive, and a square-wave lightcurve with an eclipse width
of~0.1.  The eclipse duration of this last lightcurve is $\sim 1/2$
the width of the high-state eclipse, which is what one expects if the
x-ray source was a point (rather than extended) source.

In order to determine if the observed low-state lightcurve was well
described by either the square-wave or the scaled on-state lightcurve,
we calculated the chi-squared for each of these.  The trial
lightcurves were first artificially binned to match the sampling of
the observed (7 bin) lightcurve.  For the scaled on-state lightcurve
we calculate a reduced $\chi^2$ for 6~degrees of freedom of
$\chi^2/\nu = 0.53$ (80\% random probability), and for the square wave
we find $\chi^2/\nu = 1.7$ (10\% random probability).  Clearly either
trial lightcurve is an acceptable representation of the observations,
although the scaled on-state lightcurve may be somewhat favored.

Given the limited statistics, one might reasonably ask if any source
variability has been formally detected at all.  Testing the 7~bin
lightcurve against a steady source we find $\chi^2/\nu = 1.2$, clearly
allowing that the source is steady.  However, two other tests provide
evidence that some sort of an eclipse still occurs in the
quiescent state.

First, we have cross-correlated the observed 7~bin lightcurve against
the scaled on-state curve in order to determine the phase of minimum
light.  The best fit phase and 68\% confidence limits are $0.0\pm 0.2$
on the ephemeris of McClintock \etal\/ (1982).  The accumulated error in the
ephemeris is $\pm 0.1$, so while the present data agrees with this
ephemeris it is not able to refine it.  Given the large error in the
determination of phase zero, this provides only weak evidence for an
eclipse.

Second, we binned the data into 4~equal bins, one of which is centered
on the eclipse phase.  Testing this more coarsely binned lightcurve
against a steady source we find $\chi^2/\nu = 3.3$, which has a random
probability of only $\sim 2\%$ for a steady source.  In addition, the
minimum of this binned lightcurve occurs at the expected phase of
eclipse, which has a random probability of 1 in 4.  This provides
somewhat stronger evidence that an eclipse may still be occurring at
phase zero.

\section{Discussion}

A glance at Figure~3 shows that the quiescent lightcurve appears
remarkably similar to a scaled version of the outburst lightcurve.
However, the statistics are poor, and the lightcurve binned into
7~phase intervals is formally consistent with a steady source, a
square wave (eclipse), or a scaled version of the on-state lightcurve.
However, when binned more coarsely the lightcurve provides tantalizing
evidence that there is an eclipse occurring at the
expected phase.  

This result differs somewhat from what we found in the ROSAT HRI data
obtained $\sim 2.5$~years before these data.  The HRI data excluded a
scaled version of the outburst lightcurve, and these PSPC data allow
it.  Given that the source flux decreased by a factor of $\sim 2$
between the observations, it would not be surprising if the lightcurve
also changed shape.  However, the statistical uncertainties of the HRI
lightcurve are large, and (like the PSPC lightcurve) they allow the
possibility that the source was constant.

The eclipse (square wave) lightcurve is what one expects based on the
standard models of the system.  At this low luminosity, the ADC should
have collapsed (White \& Holt 1982, Begelman, McKee \& Shields
1983), and the secondary should eclipse the x-ray emission from the
neutron star for $\sim 10\%$ of the orbit.  A smoothly modulated
lightcurve would be hard to understand, as it would imply that the
structure of the accretion disk has not changed despite the observed
large change in x-ray luminosity, and presumably mass transfer rate.

In the high state, the modulation of the x-ray flux was accompanied by
a strong modulation of the optical flux.  What optical modulation
should we expect given the apparent modulation in the low-state x-ray
flux?  In order to answer this question we need an estimate of the
quiescent disk V magnitude, which we make by comparing 4U~2129+47 to
Cen X-4.  This SXT also contains a NS primary, and the quiescent disk
has been measured to have an absolute magnitude ${\rm M_V \sim 9}$
(McClintock \& Remillard 1990).  The disk in Cen X-4 may be somewhat
bigger than that in 4U~2129+47 due to its longer period (15.1~h), and
also may appear somewhat brighter due to its more face-on viewing
angle ($i \sim 40^o$, Shahbaz, Naylor \& Charles 1993).  Based
on the amplitude of the optical modulation in outburst 
(McClintock, Remillard, \& Margon 1981)
and the correlation between modulation and inclination 
(Van Paradijs, Van der Klis, \& Pedersen 1988), the
inclination of 4U~2129+47 is likely to be $i$~\gax~80$^o$.  The
differences in the period and inclination would 
lower our estimate of the absolute magnitude of
the quiescent disk in 4U~2129+47 by several magnitudes, based
on the correlations found in cataclysmic variables (i.e., equations 2.63 and
3.3 from Warner 1995).  Thus ${\rm M_V \sim 9}$ is a comfortable lower
limit to the quiescent disk in 4U~2129+47.  
At the 6.3~kpc distance of 4U~2129+47, and with ${\rm A_V} =
1.5$, this disk would have ${\rm V} \sim 24.5$.  The quiescent counterpart
has ${\rm V}=17.9$, so even if the disk were 100\% modulated we would expect
a fractional modulation of only $\sim 0.2\%$, which is well below the
observed limit of 1.2\% amplitude (99\% confidence, Thorstensen
\etal\/ 1988).

\subsection{Comparison to Other Quiescent SXT}
	
	The x-ray temperatures of $\sim 0.2$~keV which we find for
these two SXT in quiescence are similar to those found for other
NS~SXT (Verbunt \etal\/ 1994. Asai \etal\/ 1996, Asai \etal\/ 1998
and references therein). The luminosity of 4U~2129+47 is also similar to that
found in these sources, but EXO~0748-676 is at the extreme high end of
the distribution (Narayan \etal\/ 1997a).  There is no obvious reason
for this, in particular the orbital period of EXO~0748-676 is typical,
so the average mass transfer rate should be typical as well (Menou
\etal\/ 1999).  Given 
the observed variability of quiescent NS~SXT (Campana \etal\/ 1998), it
is possible that the single quiescent measurement caught this system in
a particularly luminous state.

The emitting radius we compute assuming a black-body spectrum is
 smaller than a NS in the case of 4U~2129+47.  Black-body fits often
 indicate emitting areas smaller than a NS surface, indicating either
 that the accretion is channeled onto a small fraction of the NS
 surface (Asai \etal\/ 1996, Menou \etal\/ 1999) 
or that the
 black-body spectral fits erroneously indicate a small surface area
 (Rajagopal \& Romani 1996, Rutledge \etal\/ 1999).  EXO~0748-676 is
 unusual in that the black-body radius is consistent with the entire
 NS surface, owing to the larger than typical luminosity in
 quiescence.

In summary, we note that the difference in the luminosity swings of
outbursting and quiescent SXT may indicate the existence of event
horizons in BH~SXT (Narayan \etal\/ 1997a, Garcia \etal\/ 1998, Asai
\etal\/ 1998, Menou \etal\/ 1999, but see Chen \etal\/ 1998 for an
opposing view).  Observations with AXAF and XMM will undoubtedly add
more SXT to the studied sample, and push upper limits lower,
helping to more definitively test this difference.  XMM may allow an
accurate measurement of the quiescent light curve of the eclipsing
NS~SXT 4U~2129+47, therefore providing a spatially resolved picture of
an ADC in a quiescent system.

We thank Jeff McClintock for helpful comments on an earlier draft of
this paper. This work was supported in part by the Chanrda Science
Center through contract NAS 8-39073.

\clearpage
\newpage


\clearpage
\newpage

\begin{figure}
\psfig{figure=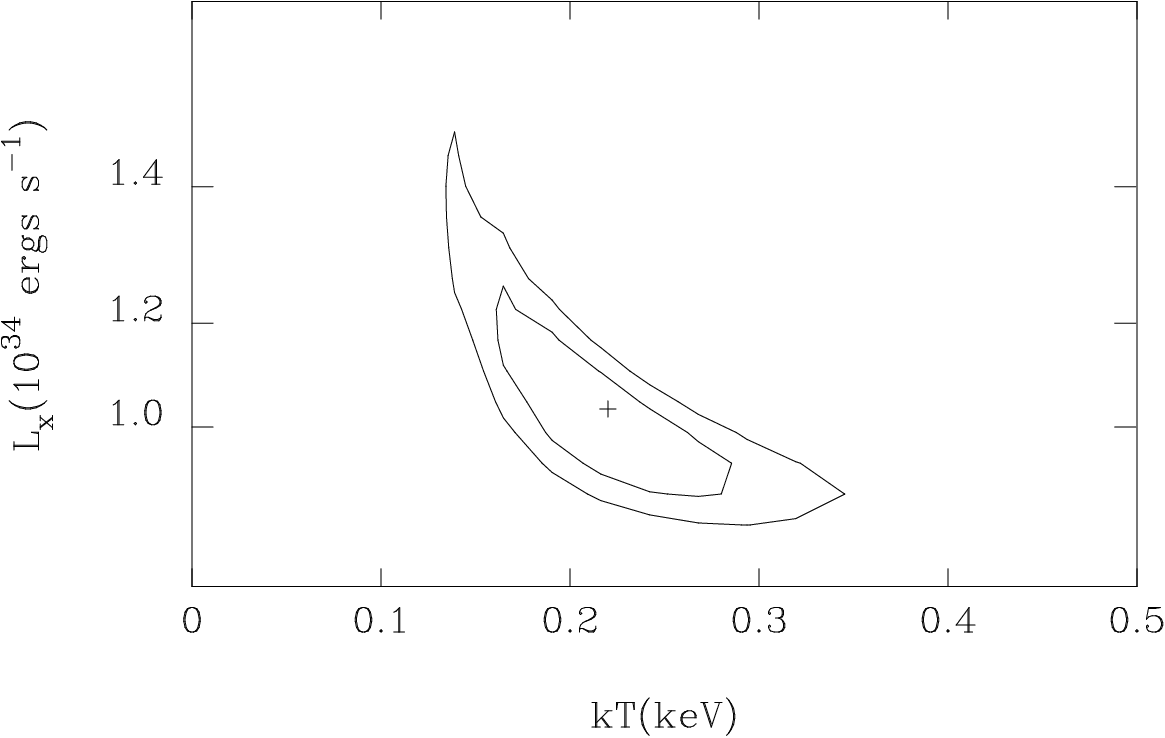,angle=-90,width=6in}
\vskip 12pt
\begin{minipage}[t]{6in}
Figure~1: The chi-squared grid for black body spectral fits
to the IPC data on EXO~0748-676 in quiescence, using Poisson
(Gehrels 1986) weighting.  The cross corresponds to the best fit
value.  The 68\% and 90\% confidence regions are shown.  The source
distance is assumed to be 10~kpc, and the absorption has been fixed at
${\rm log(N_H)} = 21.35 {\rm cm^{-2}}$, corresponding to the optical
reddening of ${\rm E(B-V)=0.42}$.
\end{minipage}
\end{figure}

\begin{figure}
\psfig{figure=MRgarcia.fig2.ps,angle=-90,width=6in}
\vskip 12pt
\begin{minipage}[t]{6in}
Figure~2: The chi-squared grid for black body spectral fits to
the PSPC data on 4U~2129+47 in quiescence, using Poisson
(Gehrels 1986) weighting.  The contours correspond to the
68\% and 90\% confidence intervals. 
The source distance is assumed to be 6.3~kpc, and the  absorption has
been fixed at ${\rm log(N_H)} = 21.2~cm^{-2}$, corresponding to the optical
extinction of ${\rm A_V} = 0.9$.  
\end{minipage}
\end{figure}

\begin{figure}
\psfig{figure=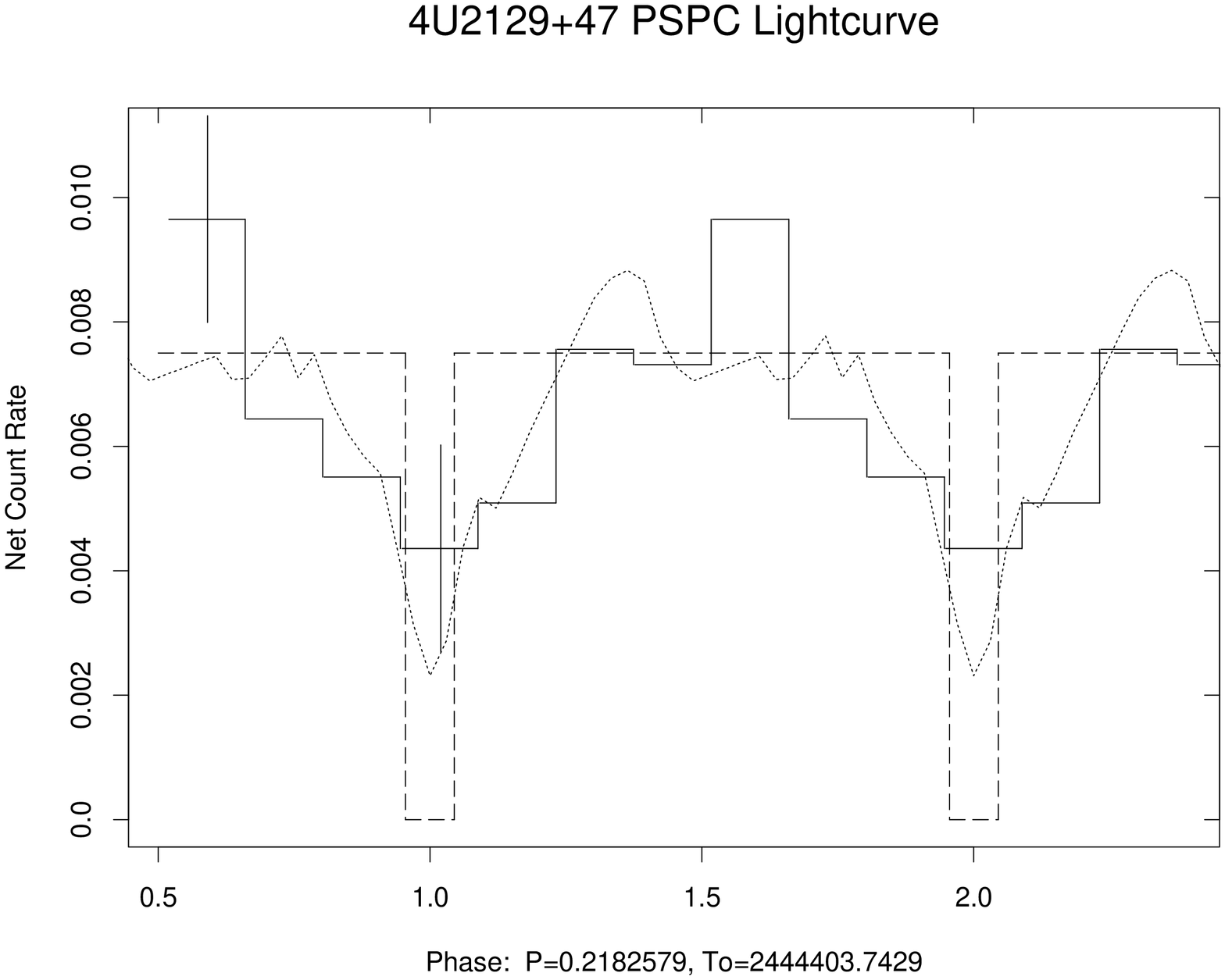,angle=-90,width=6in}
\begin{minipage}[t]{6in}
Figure~3:  The low-state orbital lightcurve for 4U~2129+47.  Three lines are shown
- the solid line is the back-ground subtracted PSPC counting rate,
the dotted line is the on-state (IPC) lightcurve scaled to match the
low-state count rate, and the dashed line shows the expected eclipse
of the neutron star by the secondary.  Either is a statically
acceptable description of the low-state lightcurve, with $\chi^2/\nu = 0.53
$ for the scaled lightcurve, and $\chi^2/\nu = 1.7$ for the eclipse
lightcurve.  
\end{minipage}
\end{figure}

\end{document}